\documentstyle[preprint,prl,aps,graphicx]{revtex}
\newtheorem{theorem}{Theorem}

\begin{document}
\draft
\title{Minimal Gauge Invariant Classes of Tree Diagrams in Gauge Theories}
\author{Edward Boos} 
\address{Institute of Nuclear Physics,
  Moscow State University, 119899, Moscow, Russia}
\author{Thorsten Ohl} 
\address{Darmstadt University of Technology,
  Schlo\ss gartenstr. 9, D-64289 Darmstadt, Germany}
\date{March 1999}
\maketitle
\begin{abstract}
  We describe the explicit construction of \emph{groves,} the smallest
  gauge invariant classes of tree Feynman diagrams in gauge theories.
  The construction is valid for gauge theories with any number of
  group factors which may be mixed. It requires no summation over a
  complete gauge group multiplet of external matter fields. The method
  is therefore suitable for defining gauge invariant classes of
  Feynman diagrams for processes with many observed final state
  particles in the standard model and its extensions.
\end{abstract}
\narrowtext
\section{Introduction}
\label{sec:introduction}

The quest for a theory of flavor demands precise calculations of high
energy scattering processes in the framework of the standard model and
its extensions.  At the Tevatron, the LHC, and at a future
$e^+e^-$~Linear Collider, final states with many detected particles
and tagged flavor will be the primary handle for testing theories of
flavor.  Calculations of cross section with many particle final states
remain challenging and it is of crucial importance to be able to
concentrate on the important parts of the scattering amplitude for the
phenomena under consideration.

In gauge theories, however, it is impossible to simply select the
signal diagrams and to ignore irreducible backgrounds.  The same
subtle cancellations among the diagrams in a gauge invariant subset
that lead to the celebrated good high energy behavior of gauge
theories such as the standard model, come back to
haunt us if we accidentally select a subset of diagrams that is not
gauge invariant.  The results from such a calculation have \emph{no}
predictive power, because they depend on unphysical parameters
introduced during the gauge fixing of the Lagrangian.  It must be
stressed that not all diagrams in a gauge invariant subset have the
same pole structure and that a selection based on `signal' or
`background' will not suffice.

The subsets of Feynman diagrams selected for any calculation must
therefore form a \emph{gauge invariant subset}, i.\,e.~together they
must already satisfy the Ward and Slavnov-Taylor identities to insure
the cancellation of contributions from unphysical degrees of freedom.

In abelian gauge theories, such as QED, the classification of gauge
invariant subsets is straightforward and can be summarized by the
requirement of inserting any additional photon into \emph{all}
connected charged
propagators.  This situation is similar for gauge theories with simple
gauge groups, the difference being that the gauge bosons are carrying
charge themselves.  For non-simple gauge groups like the standard
model, which even includes mixing, the classification of gauge
invariant subsets is much more involved.

Indeed, up to now, the classification of gauge invariant subsets in
the standard model has been performed in an ad-hoc fashion
(cf.~\cite{Bardin/etal:1994:4f-classification,Boos/Ohl:1997:gg4f}).
In this note we present an explicit construction of \emph{groves,} the
smallest gauge invariant classes of tree Feynman diagrams in gauge
theories.  Our construction is not restricted to gauge theories with
simple gauge groups.  Instead, it is applicable to gauge groups with
any number of factors, which can even be mixed, as in the standard
model.  Furthermore, it does not require a summation over complete
multiplets and can therefore be used in flavor physics when members of
weak isospin doublets (such as charm or bottom) are detected
separately.  Our method constructs the smallest gauge
invariant subsets.  Below we show examples in which they are indeed
smaller than those derived from looking at final state
alone~\cite{Bardin/etal:1994:4f-classification,Boos/Ohl:1997:gg4f}.

We expect that our methods will also have applications in loop
calculations.  However, some of our current proofs use properties of
tree diagrams and further research is required in this area.

\section{Forests}
\label{sec:forests}

\begin{figure}
  \begin{center}
    \includegraphics{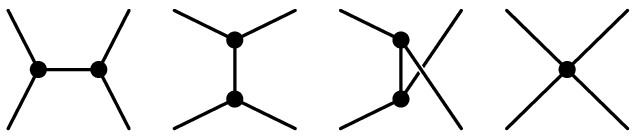}
  \end{center}
  \caption{\label{fig:flips}%
    The four-point diagrams $\{S_1,S_2,S_3,S_4\}$ related by
    \emph{flips}.}
\end{figure}

We introduce basic notions in the case of unflavored scalar $\phi^3$-
and $\phi^4$-theory.  In the absence of selection rules, the
diagrams~$S_1$, $S_2$, and~$S_3$ in figure~\ref{fig:flips} must have
the same coupling strength to ensure crossing invariance.
If there are additional symmetries, as in the case of gauge theories,
the coupling of~$S_4$ will be fixed relative to $S_{1,2,3}$.

The elementary \emph{flips}~$S_{t_4}\to S_{t'_4}$ define
relations~$t_4\circ t'_4$ on the set~$T_4$ of all tree graphs with
four external particles.  These trivial relations have a non-trivial
natural extension to the set~$T(E)$ of \emph{all} tree diagrams with a
given external state~$E$ by
\begin{equation}
\label{eq:extend-flip}
   t\circ t' \Longleftrightarrow
     \exists t_4\in T_4,t'_4\in T_4:
       t_4 \circ t'_4 \land t\setminus t_4 = t'\setminus t'_4,
\end{equation}
i.\,e.~two diagrams satisfy the relation if they are identical up to a
single flip of a four-point subdiagram.  This relation allows to view
the set~$T(E)$ of all tree diagrams as the vertices of a graph~$F(E)$
\begin{equation}
\label{eq:forest}
  F(E) = \bigl\{(t,t')\in T(E)\times T(E) \bigl| t\circ t'\bigl\},
\end{equation}
where the edges of the graph are formed by the pairs of diagrams
related by a single flip.  To avoid confusion, we will refer to
graph~$F(E)$ as \emph{forest} and to its vertices as Feynman
\emph{diagrams}.  For lack of space, we have to introduce some
mathematical concepts rather tersely and will give a more
self-contained presentation elsewhere~\cite{Boos/Ohl:1999:proof}.

Already the simplest non-trivial example of such a forest, the 15~tree
diagrams with five external particles in unflavored $\phi^3$-theory,
as shown in figure~\ref{fig:5phi3}, displays an intriguing structure.
The most important property for our applications is
\begin{theorem}
\label{th:forest}
  The unflavored forest~$F(E)$ is connected for all external states~$E$,
\end{theorem}
which is easily proved by induction on the number of particles
in the external state.  This theorem shows that it is possible to
construct \emph{all} Feynman diagrams by visiting the nodes of~$F(E)$
along successive applications of the flips in figure~\ref{fig:flips}.

\section{Flavored Forests}
\label{sec:flavor}

In physics applications we have to deal with different particles.
Therefore we introduce \emph{flavored forests}, where the
admissibility of elementary flips~$t_4\circ t_4'$ depends on the
four particles involved through the Feynman rules for the vertices
in~$t_4$ and~$t_4'$. Flavored forests have in general more than one
connected component.

In order to simplify the combinatorics when counting diagrams for
theories with more than one flavor, we will below treat \emph{all}
external particles as outgoing.  The physical amplitudes are obtained
later by selecting the incoming particles in the end.  Ward
identities, etc.~will be proved for the latter physical amplitudes, of
course.

\section{Groves}
\label{sec:groves}

\begin{figure}
  \begin{center}
    \includegraphics{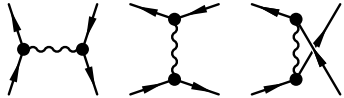}
  \end{center}
  \caption{\label{fig:F}%
    The four-point diagrams $\{t_4^{F,1},t_4^{F,2},t_4^{F,3}\}$
    related by \emph{flavor flips}.}
\end{figure}

\begin{figure}
  \begin{center}
    \includegraphics{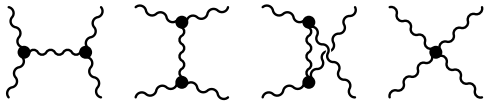}
  \end{center}
  \caption{\label{fig:G}%
    The four-point diagrams
    $\{t_4^{G,1},t_4^{G,2},t_4^{G,3},t_4^{G,4}\}$ related by
    \emph{gauge flips}.}
\end{figure}

\begin{figure}
  \begin{center}
    \includegraphics{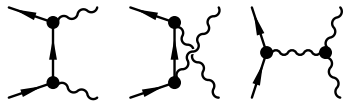}
  \end{center}
  \caption{\label{fig:C}%
    The four-point diagrams
    $\{t_4^{G,5},t_4^{G,6},t_4^{G,7}\}$ related by
    \emph{gauge flips} in the case of fermionic matter.}
\end{figure}

Our method is based on the observation that the flips in gauge
theories fall into two different classes: the \emph{flavor flips} in
figure~\ref{fig:F} which involve four matter fields which carry gauge
charge and possibly additional conserved quantum numbers
and the \emph{gauge flips} in figures~\ref{fig:G} and~\ref{fig:C} which
also involve gauge bosons
(another diagram, $t_4^{G,8}\propto \phi^2A^2$, has to be added for
scalar matter fields that appear in extensions of the standard model:
SUSY partners, leptoquarks, etc.).  In gauge theories with more than
one factor, like the standard model, the gauge flips are extended in
the obvious way to include all four-point functions with at least one
gauge-boson. Commuting gauge group factors lead to separate sets, of
course.  In spontaneously broken gauge theories, the Higgs and
Goldstone boson fields contribute additional flips, in which they are
treated like gauge bosons (see~\cite{Boos/Ohl:1999:proof} for a
complete list and applications). Ghosts can be ignored at tree level.

The flavor flips~(figure~\ref{fig:F}) are special because they can be
switched 
off without spoiling gauge invariance by introducing a horizontal
symmetry that commutes with the gauge group.  Such a horizontal
symmetry is similar to the replicated generations in the standard
model, but if three generations do not suffice, it can also be introduced
artificially.  Typical examples are Bhabha-scattering, where the
$s$-channel and the $t$-channel diagrams are separately gauge
invariant, because we can replace one electron line by a muon line without
violating gauge invariance.  Similarly, the charged current and
neutral current contributions in~$ud\to ud$ can be switched on and off by
assuming that two of the four quarks are from a different generation.

This observation suggests to introduce two relations:
\begin{equation}
  t\bullet t'  \Longleftrightarrow
    \text{$t$ and~$t'$ related by a \emph{gauge} flip}
\end{equation}
and
\begin{equation}
  t\circ t' \Longleftrightarrow
    \text{$t$ and~$t'$ related by a \emph{flavor or gauge} flip}.
\end{equation}
These two relations define two different graphs with the same set~$T(E)$ of
all Feynman diagrams as vertices:
\begin{equation}
  F(E) = \bigl\{(t,t')\in T(E)\times T(E) \bigl| t\circ t'\bigl\}
\end{equation}
and
\begin{equation}
  G(E) = \bigl\{(t,t')\in T(E)\times T(E) \bigl| t\bullet t'\bigr\}.
\end{equation}
For brevity, we will continue to denote the \emph{flavor
forest}~$F(E)$ as the \emph{forest} of the external state~$E$ and we
will denote the connected components~$G_i(E)$ of the \emph{gauge
forest}~$G(E)$ as the \emph{groves} of~$E$.
Since~$t\bullet t'\Rightarrow t\circ t'$, we
have~$\bigcup_i G_i(E)=G(E)\subseteq F(E)$, i.\,e.~the groves are a
\emph{partition} of the forest.

\begin{theorem}
\label{th:groves}
  The forest~$F(E)$ is connected if the fields in~$E$ carry no
  conserved quantum numbers other than the gauge charges.
  The groves~$G_i(E)$ are the minimal gauge invariant classes of
  Feynman diagrams.
\end{theorem}
Here we give a sketch of the proof, which will be presented in more
detail elsewhere~\cite{Boos/Ohl:1999:proof}.  As we have seen, the
theorem is true for the four-point diagrams and we can use induction
on the number of external matter fields and gauge bosons. Since the
matter fields are carrying conserved charges, they can only be added
in pairs.

If we add an additional external gauge boson to a gauge invariant
amplitude, the diagrammatical proof of the Ward and Slavnov-Taylor
identities in gauge theories requires us to sum over all ways to
attach a gauge boson to connected gauge charge carrying components of
the Feynman diagrams.  However, the gauge flips are connecting pairs of 
neighboring insertions and can be iterated along gauge charge carrying
propagators.  Therefore no partition of the forest~$F(E)$ that is
finer than the groves~$G_i(E)$ preserves gauge invariance.

\begin{figure}
  \begin{center}
    \includegraphics{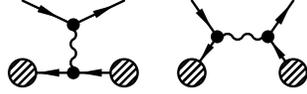}
  \end{center}
  \caption{\label{fig:matter-pair}%
    The two ways of adding a matter field pair to a diagram.}
\end{figure}

If we add an additional pair of matter fields to a gauge invariant
amplitude, we have to consider two separate cases, as shown in
figure~\ref{fig:matter-pair}. If the new flavor
does not already appear among the other matter fields, the only way
to attach the pair is through a gauge boson.  If the new flavor is
already present, we can also break up a matter field propagator and
connect the two parts of the diagram with a new gauge propagator.
Since it is always possible to introduce a new flavor, either physical
or fictitious, without breaking gauge invariance, these cases
fork off separately gauge invariant classes every time we add a new
pair of matter fields.  On the other hand, the cases
in figure~\ref{fig:matter-pair} are related by a flavor flip.
Therefore~$F(E)$ remains connected, the~$G_i(E)$ are separately gauge
invariant and the proof is complete.

Earlier
attempts~\cite{Bardin/etal:1994:4f-classification,Boos/Ohl:1997:gg4f}
have used physical final states as a criterion for identifying gauge
invariant subsets.  We have already shown that the groves are minimal
and therefore never form a more coarse partition than the one derived
from a consideration of the final states alone.  Below we shall see
examples where the groves do indeed provide a strictly finer partition.

In a practical application one calculates the groves for the
interesting combinations of gauge quantum numbers, such weak
isospin and hypercharge in the standard model, using an external state
where all other quantum numbers are equal.  The physical amplitude
is then obtained by selecting the groves that are compatible with the
other quantum numbers of the process under consideration.  Concrete
examples are considered in the next section.

\section{Application}
\label{sec:application}

In table~\ref{tab:6f}, we list the groves for all processes with six
external massless fermions in the standard model, with and without
single photon bremsstrahlung, without QCD and CKM mixing.  We can
easily include fermion masses, QCD, and CKM mixing within the same
formalism, but the table would have to be much larger, because
additional gluon, Higgs and Goldstone
diagrams appear, depending on whether the fermions are massive or
massless, colored or uncolored. In the table, cases with
identical~$\mathrm{SU}(2)_L\otimes \mathrm{U}(1)_Y$ quantum numbers
are listed only once and cases with different~$T_3$ and~$Y$ are listed
separately only if the vanishing of the electric charge
removes diagrams from a grove.

The familiar non-minimal gauge invariant classes for $e^+e^-\to
f_1\bar f_2 f_3\bar f_4$ \cite{Bardin/etal:1994:4f-classification} are
included in table~\ref{tab:6f} as special cases.  The LEP2
$WW$-classes CC09, CC10, and~CC11 are immediately obvious.  As a not
quite so obvious example, the
process~$e^+e^-\to\mu^+\mu^-\tau^+\tau^-$ has the same
$\mathrm{SU}(2)_L$~quantum numbers as~$u\bar u\to u\bar uu\bar u$.  We
can read off table~\ref{tab:6f} that, in the case of identical pairs,
there are 18~groves, of 8~diagrams each.  If all three pairs are
different, the number of groves has to be divided by~$3!$, because we
are no longer free to connect the three particle-antiparticle pairs
arbitrarily.  Thus there are 24~diagrams contributing to the
process~$e^+e^-\to\mu^+\mu^-\tau^+\tau^-$ and they are organized in
three groves of 8~diagrams each.  Any diagram in a grove can be
reached from the other 7 by exchanging the vertices of the gauge bosons
on one fermion line and be exchanging~$Z^0$ and~$\gamma$.
Since there are no non-abelian 
couplings in this process, the separate gauge invariance of each grove
could also be proven as in QED, by varying the hypercharge of each
particle: $A\propto A_1\cdot q_e^2 q_\mu q_\tau + A_2\cdot q_e q_\mu^2
q_\tau + A_3 \cdot q_e q_\mu q_\tau^2$.  

In figure~\ref{fig:mix71} with show the forest for the
process~$\gamma\gamma\to u\bar d d\bar u$ in the standard model.
The grove in the center consists of the
31~diagrams with charged current interactions (the set CC21
of~\cite{Boos/Ohl:1997:gg4f}).  The four small groves of neutral
current interactions are only connected with the rest of the forest
through the charged current grove.

The groves can now be used to select the Feynman diagrams to be
calculated by other means.  However, we note that it is also possible
to calculate the amplitude with little additional effort already
during the construction of the groves by keeping track of momenta and
couplings in the diagram flips.

\section{Automorphisms}
\label{sec:automorphisms}

The forest and groves that we have studied appear to be very
symmetrical in the neighborhood of any vertex.  However, the global
connection of these neighborhoods is twisted, which makes it all but
impossible to draw the graphs in a way that makes these apparent
symmetries manifest. 

\begin{figure}
  \begin{center}
    \includegraphics{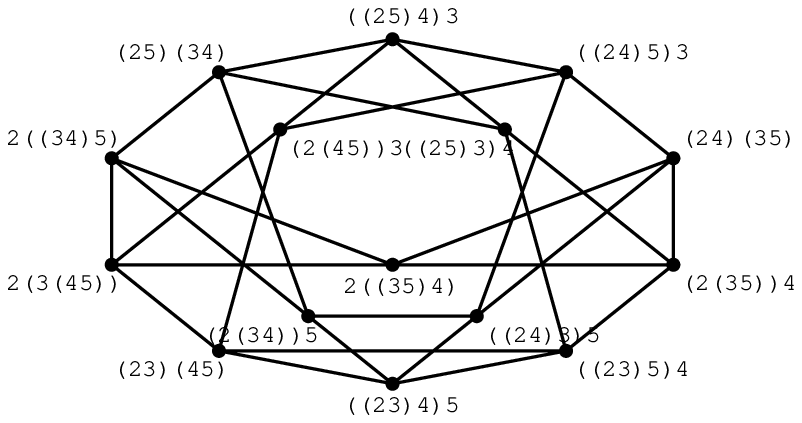}
  \end{center}
  \caption{\label{fig:5phi3}%
    The forest of the 15 five-point tree diagrams in unflavored
    $\phi^3$-theory. The diagrams are specified by fixing vertex~1 and
    using parentheses to denote the order in which lines are joined at
    vertices.}
\end{figure}

\begin{figure}
  \begin{center}
    \includegraphics[width=.45\textwidth]{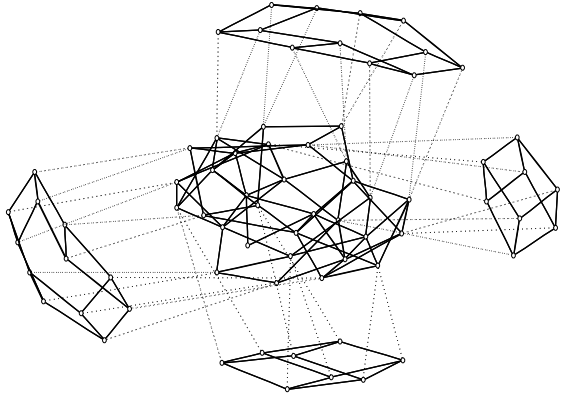}
  \end{center}
  \caption{\label{fig:mix71}%
    The forest of size~71 for the process~$\gamma\gamma\to u\bar d d\bar u$ in
    the standard model (without QCD, CKM mixing and Higgs
    contributions and in unitarity gauge) with one grove of size~31,
    two groves of size 12 and two groves of size 8. Solid lines represent gauge flips and
    dotted lines represent flavor flips.}
\end{figure}

Nevertheless, one can turn to mathematics~\cite{McKay:1990:nauty} and
construct the automorphism groups $\text{\textbf{Aut}}(F(E))$ and
$\text{\textbf{Aut}}(G_i(E))$ of the
forest~$F(E)$ and the groves~$G_i(E)$, i.\,e.~the group of
permutations of vertices that leave the edges invariant.  These groups
turn out to be larger than one might expect.  For example, the group
of permutations of the 71~vertices of the forest~$F(\gamma\gamma\to
u\bar d d\bar u)$ in figure~\ref{fig:mix71}, that leave the edges
invariant, has 128~elements.  Similarly, the automorphism group of the
forest in figure~\ref{fig:5phi3} has $120=5!$~elements.

The study of these groups and their relations might enable us to
construct gauge invariant subsets directly.  This is however
beyond the scope of the present note and will be considered elsewhere.
 
\paragraph*{Acknowledgments.}
We thank Alexander Pukhov for useful discussions.  This work was
supported in part by Deutsche Forschungsgemeinschaft (MA\,676/5-1).
E.\,B.~is grateful to the Russian Ministry of Science and
Technologies, and to the Sankt-Petersburg Grant Center
for partial financial support. T.\,O.~is supported by
Bundesministerium f\"ur Bildung, Wissenschaft, Forschung und
Technologie, Germany (05\,7SI79P\,6, 05\,HT9RDA).



\begin{table}
  \caption{\label{tab:6f}%
    The groves for all processes with six external massless fermions
    and single photon bremsstrahlung in the standard model (without
    QCD). Six-fermion processes with identical groves can have different
    groves if a photon is added, because the photon distinguishes
    leptons from neutrinos in internal lines.}
  \begin{center}
    \begin{tabular}{lrl}
      external fields ($E$) & diagrams & classes \\\hline
      $u \bar u u \bar u u \bar u$
        & $144$ & $18\cdot8$ \\
      $u \bar u u \bar u u \bar u\gamma$
        & $1008$ & $18\cdot24+36\cdot16$ \\
      $u \bar u u \bar u d \bar d$
        & $92$ & $4\cdot11+6\cdot8$ \\
      $u \bar u u \bar u d \bar d\gamma$
        & $696$ & $4\cdot90+6\cdot24+12\cdot16$ \\
      $\ell^+ \ell^- u \bar u d \bar d$
        & $27$ & $1\cdot11+2\cdot8$ \\
      $\ell^+ \ell^- u \bar u d \bar d\gamma$
        & $201$ & $1\cdot89+2\cdot24+4\cdot16$ \\
      $\ell^- \nu d \bar u d \bar d$
        & $20$ & $2\cdot10$\\
      $\ell^- \nu d \bar u d \bar d\gamma$
        & $142$ & $2\cdot71$ \\
      $\ell^+ \ell^- \ell^- \nu d \bar u$
        & $20$ & $2\cdot10$\\
      $\ell^+ \ell^- \ell^- \nu d \bar u\gamma$
        & $140$ & $2\cdot70$ \\
      $\ell^- \nu \ell^+ \bar\nu d \bar d$
        & $15$ & $1\cdot9+1\cdot4+1\cdot2$ \\
      $\ell^- \nu \ell^+ \bar\nu d \bar d\gamma$
        & $82$ & $1\cdot54+1\cdot12+1\cdot8+2\cdot4$ \\
      $\ell^- \bar\nu \ell^+ \nu \ell^+ \ell^-$
        & $56$ & $4\cdot9+4\cdot4+2\cdot2$ \\
      $\ell^- \bar\nu \ell^+ \nu \ell^+ \ell^-\gamma$
        & $308$ & $4\cdot53+4\cdot12+4\cdot8+4\cdot4$ \\
      $\ell^+ \nu \ell^- \bar\nu \nu \bar\nu$
        & $36$ & $4\cdot6+6\cdot2$ \\
      $\ell^+ \nu \ell^- \bar\nu \nu \bar\nu\gamma$
        & $120$ & $4\cdot23+2\cdot6+4\cdot4$ \\
      $\nu \bar\nu \nu \bar\nu \nu \bar\nu$
        & $36$ & $18\cdot2$
    \end{tabular}
  \end{center}
\end{table}

\end{document}